\documentclass[twocolumn,showpacs,preprintnumbers,amsmath,amssymb,aps]{revtex4}

\usepackage{graphicx}
\usepackage{dcolumn}
\usepackage{bm}

\begin{document}

\title{Complete physical simulation of the entangling-probe  attack on the BB84 protocol}

\author{Taehyun Kim\footnote{Electronic address: thkim@mit.edu}, Ingo Stork genannt Wersborg, Franco N. C. Wong, and Jeffrey H. Shapiro}
 \affiliation{Research Laboratory of Electronics, Massachusetts Institute of 
Technology, Cambridge, Massachusetts 02139, USA}

\date{\today}

\begin{abstract}
We have used deterministic single-photon two qubit (SPTQ) quantum logic to implement the most powerful individual-photon attack against the Bennett-Brassard 1984 (BB84) quantum key distribution protocol.  Our measurement results, including physical source and gate errors, are in good agreement with theoretical predictions  for the R{\'e}nyi information obtained by Eve as a function of  the errors she imparts to Alice and Bob's sifted key bits.  The current experiment is a physical simulation of a true attack, because Eve has access to Bob's physical receiver module.  This experiment illustrates the utility of an efficient deterministic quantum logic for performing realistic physical simulations of quantum information processing functions.  

\end{abstract}

\pacs{03.67.Dd, 03.67.-a, 42.65.Lm}

\maketitle
In 1984 Bennett and Brassard \cite{BB84} proposed a protocol (BB84) for quantum key distribution (QKD) in which the sender (Alice) transmits single-photon pulses to the receiver (Bob) in such a way that security is vouchsafed by physical laws.  Since then, BB84 has been the subject of many security analyses \cite{QKDreview}, particularly for configurations that involve nonideal operating conditions, such as the use of weak laser pulses in lieu of single photons.  A more fundamental question is how much  information the eavesdropper (Eve) can gain under ideal BB84 operating conditions. Papers by Fuchs and Peres \cite{FP}, Slutsky {\it et al}.\ \cite{Slutsky}, and Brandt \cite{FPB} show that the most powerful individual-photon attack can be accomplished with a controlled-{\sc not} ({\sc cnot}) gate.   As illustrated in Fig.~1, Eve supplies the target qubit to the {\sc cnot} gate, which entangles it with the BB84 qubit that Alice is sending to Bob.  Eve then makes her measurement of the target qubit to obtain information on the shared key bit at the expense of imposing detectable errors between Alice and Bob \cite{FPB,SIM}.   

\begin{figure}[ht]
\center{\includegraphics[width=0.9\linewidth]{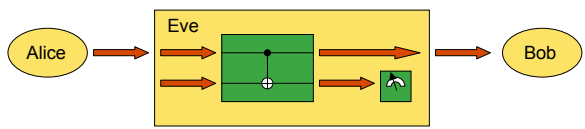}}
\caption{(Color online) Block diagram of the Fuchs-Peres-Brandt probe for attacking BB84 QKD.}
\label{FPB_probe}
\end{figure} 

We have recently shown \cite{SIM} that this Fuchs-Peres-Brandt (FPB) entangling probe can be implemented using single-photon two-qubit (SPTQ) quantum logic in a 
proof-of-principle experiment.  In SPTQ logic a single photon carries two independent qubits: the polarization and the momentum (or spatial orientation) states of the photon.  Compared to standard two-photon quantum gates, SPTQ gates are deterministic and can be efficiently implemented using only linear optical elements.  We have previously demonstrated {\sc cnot} \cite{SPTQ} and {\sc swap} \cite{SWAP} gates in this SPTQ quantum logic platform.  SPTQ logic affords a simple yet powerful way to investigate few-qubit quantum information processing tasks.  

In this work we use SPTQ logic to implement the FPB probe as a complete physical simulation of the most powerful attack on BB84, including physical errors.  This is to our knowledge the first experiment on attacking BB84, and the results are in good agreement with theoretical predictions.  It is only a physical simulation because the two qubits of a single photon carrier must be measured jointly, so that Eve needs access to Bob's  receiver, but \em not\/\rm\ his measurement.  The SPTQ probe could become a true attack if quantum nondemolition measurements were available to Eve \cite{SIM}.  Nevertheless, the physical simulation allows investigation of the fundamental security limit of BB84 against eavesdropping in the presence of realistic physical errors, and it affords the opportunity to study the effectiveness of privacy amplification when BB84 is attacked.

\begin{figure}[ht]
\center{\includegraphics[width=0.9\linewidth]{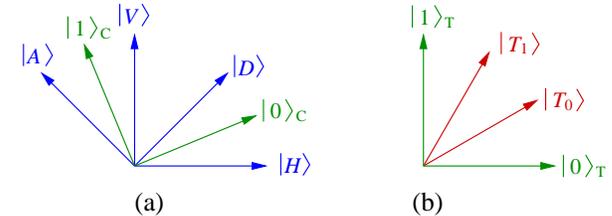}}
\caption{(Color online) Relations between different bases. (a) Control qubit basis for Eve's {\sc cnot} gate referenced to the BB84 polarization states. (b) $|T_0\rangle$ and $|T_1\rangle$ relative to the target qubit basis.}
\label{Basis}
\end{figure} 

In BB84, Alice sends Bob a single photon randomly chosen from the four polarization states of the horizontal-vertical ($H$--$V$) and $\pm 45^{\circ}$ 
($D$--$A$) bases.  In the FPB attack, Eve sets up  her {\sc cnot} gate  with its 
control-qubit computational basis, $\{|0\rangle_{\rm C},|1\rangle_{\rm C}\}$, given by $\pi/8$ rotation from the BB84 $H$-$V$ basis, as shown in Fig.~\ref{Basis}(a),
\begin{eqnarray}
|0\rangle_{\rm C} = & \cos(\pi/8)|H\rangle+\sin(\pi/8)|V\rangle\,, \nonumber \\
|1\rangle_{\rm C} = & -\sin(\pi/8)|H\rangle+\cos(\pi/8)|V\rangle\,. \label{EQ:BASIS}
\end{eqnarray}
Having selected the error probability, $P_E$, that she is willing to create, Eve prepares her probe qubit ({\sc cnot}'s target) in the initial state
\begin{eqnarray}
|T_{\rm in}\rangle & = & \{(C+S)|0\rangle_{\rm T}+(C-S)|1\rangle_{\rm T}\}/\sqrt{2}\nonumber\\
                  & \equiv & \cos\theta_{\rm in}|0\rangle_{\rm T}+\sin\theta_{\rm in}|1\rangle_{\rm T}\label{eq:T_in}
\end{eqnarray}
where $C=\sqrt{1-2P_E}, S=\sqrt{2P_E}$, and $\{|0\rangle_{\rm T}, |1\rangle_{\rm T}\}$ is the target qubit's computational basis.  After the {\sc cnot} operation---with inputs from Alice's photon and Eve's probe---the two qubits become entangled.  For each of Alice's four possible inputs, $|H\rangle$, $|V\rangle$, $|D\rangle$, and $|A\rangle$, the output of the {\sc cnot} gate is 
\begin{eqnarray}
|H\rangle|T_{\rm in}\rangle & \rightarrow & |H_{\rm out}\rangle  \equiv |H\rangle|T_0\rangle + |V\rangle|T_E\rangle\,, \label{eq:HT}\\
|V\rangle|T_{\rm in}\rangle & \rightarrow & |V_{\rm out}\rangle \equiv |V\rangle|T_1\rangle + |H\rangle|T_E\rangle\,, \label{eq:VT}\\
|D\rangle|T_{\rm in}\rangle & \rightarrow & |D_{\rm out}\rangle \equiv |D\rangle|T_0\rangle - |A\rangle|T_E\rangle\,, \label{eq:PT}\\
|A\rangle|T_{\rm in}\rangle & \rightarrow & |A_{\rm out}\rangle \equiv |A\rangle|T_1\rangle - |D\rangle|T_E\rangle\,, \label{eq:MT}
\end{eqnarray}
where $|T_0\rangle$, $T_1\rangle$, and $|T_E\rangle$ are defined in the target qubit's computational basis (see Fig.~\ref{Basis}(b)) as:
\begin{eqnarray}
|T_{0}\rangle & \equiv & \left(\frac{C}{\sqrt{2}}+\frac{S}{2}\right)|0\rangle_{\rm T} + \left(\frac{C}{\sqrt{2}}-\frac{S}{2}\right)|1\rangle_{\rm T}\,, \label{eq:T_0}\\
|T_{1}\rangle & \equiv & 
\left(\frac{C}{\sqrt{2}}-\frac{S}{2}\right)|0\rangle_{\rm T} + \left(\frac{C}{\sqrt{2}}+\frac{S}{2}\right)|1\rangle_{\rm T}\,, \label{eq:T_1}\\
|T_E\rangle & \equiv & \frac{S}{2} \left(|0\rangle_{\rm T}-|1\rangle_{\rm T}\right)\,.
\end{eqnarray}

Consider the case in which Bob measures in the same basis that Alice employed and his outcome matches what Alice sent.  Then, according to 
Eqs.~(\ref{eq:HT})--(\ref{eq:MT}), the target qubit is projected into either $|T_0\rangle$ or $|T_1\rangle$.  After Alice and Bob compare their basis selections over the classical channel, Eve can learn about their shared bit value by distinguishing between the $|T_0\rangle$ and $|T_1\rangle$ output states of her target qubit. To do so, she employs the minimum error probability receiver for distinguishing between $|T_0\rangle$ and $|T_1\rangle$ by  performing a projective measurement along $|0\rangle_{\rm T}$ and $|1\rangle_{\rm T}$.  Eve can then correlate the measurement of $|0\rangle_{\rm T}$ ($|1\rangle_{\rm T}$) with $|T_0\rangle$ ($|T_1\rangle$).  Note that this projective measurement is not perfect unless $|T_0\rangle$ and $|T_1\rangle$ are orthogonal and hence coincide with the target's computational basis, $|0\rangle_{\rm T}$ and $|1\rangle_{\rm T}$.

\begin{figure}[ht]
\center{\includegraphics[width=1\linewidth]{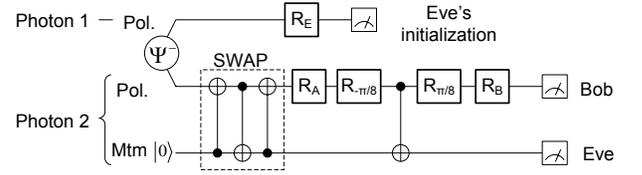}}
\caption{Quantum circuit diagram for the FPB-probe attack. Photon 1 of a 
polarization-entangled singlet photon pair heralds photon 2 and sets Eve's probe qubit to its initial state. The {\sc swap} gate allows Alice's qubit to be set in the polarization mode of photon 2, whose momentum mode is Eve's probe qubit.  The {\sc cnot} gate entangles Alice's qubit with Eve's qubit.  ${\rm R_E}$, rotation by Eve; ${\rm R_A}$, rotation by Alice; ${\rm R_B}$, rotation by Bob; ${\rm R_{\pm\pi/8}}$, rotation by angle $\pm\pi/8$.}
\label{QUANTUM_CIRCUIT}
\end{figure}

Of course, Eve's information gain comes at a cost: Eve has caused an error event whenever Alice and Bob choose a common basis and Eve's probe output state is $|T_E\rangle$.  When Alice sent $|H\rangle$ and Bob measured in the $H$--$V$ basis, Eq.~(\ref{eq:HT}) then shows that Alice and Bob will have an error event if the measured output state is $|V\rangle|T_E\rangle$. The probability that this will occur is $\langle T_E|T_E \rangle=S^2/2=P_E$. For the other three cases in 
Eqs.~(\ref{eq:VT})--(\ref{eq:MT}), the error event corresponds to the last term in each expression. Therefore the  conditional error probabilities are identical, and hence $P_E$ is the unconditional error probability.

\begin{figure}[ht]
\center{\includegraphics[width=0.9\linewidth]{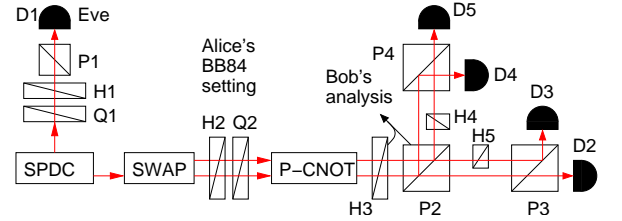}}
\caption{(Color online) Experimental configuration for a complete physical simulation of the FPB attack on BB84. SPDC, spontaneous parametric down-conversion source; H, half-wave plate; Q, quarter-wave plate; P, polarizing beam splitter; D, single-photon detector.}
\label{EXPSETUP}
\end{figure} 

To quantify Eve's information gain, we use the R\'{e}nyi information that she derives about the sift events in which Bob correctly measures Alice's qubit. Let $B=\{0,1\}$ and $E=\{0,1\}$ denote the ensembles of possible bit values that Bob and Eve receive on an error-free sift event. The R\'{e}nyi information (in bits) that Eve learns about each 
error-free sift event is
\begin{equation}
I_R \equiv -\log_2\left(\sum_{b=0}^{1}P^2(b)\right)+\sum_{e=0}^{1} P(e) \log_2 \left( \sum_{b=0}^{1}P^2(b|e)\right), \label{def_I_R}
\end{equation}
where $\{P(b),P(e)\}$ are the {\it a priori} probabilities for Bob's and Eve's bit values, and $P(b|e)$ is the conditional probability for Bob's bit value to be $b$ given that Eve's is $e$. According to Bayes' rule, $P(b|e)$ can be calculated in terms of $P(b,e)$ which is the probability of both Bob's bit to be $b$ and Eve's bit to be $e$:
\begin{equation}
P(b|e)=P(b,e)/\sum_{b=0}^1 P(b,e),
\end{equation}
where $P(b=0,e)=|(\langle H|\otimes{_{\rm T}\langle e|})|H_{\rm out}\rangle|^2$ and $P(b=1,e)=|(\langle V|\otimes{_{\rm T}\langle e|})|V_{\rm out}\rangle|^2$ in the case of $H$--$V$ basis ($|H\rangle$ corresponds to $b=0$).   A similar calculation for the  
$D$--$A$ basis, with $|D\rangle$ corresponding to $b=0$ then leads to the final result
\begin{equation}
I_R=\log_2 \left(1+\frac{4P_E(1-2P_E)}{(1-P_E)^2}\right).\label{eq:I_R}
\end{equation}
Ideally, Eve's R\'{e}nyi information is the same for both the $H$--$V$ and $D$--$A$ bases, but in actual experiments it may differ, owing to differing equipment errors in each basis.

\begin{table*}[tb]
\begin{tabular}{c r r r r r r r r r r r r r r r r}
\hline\hline
& & &\multicolumn{4}{c}{Coincidence} & &\multicolumn{4}{c}{Estimated} & &\multicolumn{4}{c}{Expected}\\ 
\cline{4-7} \cline{9-12} \cline{14-17}
Alice\hspace*{0.1in}&\multicolumn{1}{c}{$P_E$} 
& & \multicolumn{1}{c}{\hspace{.1in}$|1\rangle|0\rangle$} & \multicolumn{1}{c}{$|1\rangle|1\rangle$} 
&\multicolumn{1}{c}{$|0\rangle|1\rangle$} & \multicolumn{1}{c}{$|0\rangle|0\rangle$} 
& & \multicolumn{1}{c}{\hspace{.1in}$|1\rangle|0\rangle$} & \multicolumn{1}{c}{$|1\rangle|1\rangle$} 
&\multicolumn{1}{c}{$|0\rangle|1\rangle$} & \multicolumn{1}{c}{$|0\rangle|0\rangle$} 
& & \multicolumn{1}{c}{\hspace{.1in}$|1\rangle|0\rangle$} & \multicolumn{1}{c}{$|1\rangle|1\rangle$} 
&\multicolumn{1}{c}{$|0\rangle|1\rangle$} & \multicolumn{1}{c}{$|0\rangle|0\rangle$} \\ \hline

 & $0$ & & 1356 & 1836 & 23408 & 23356 & & 0.027 & 0.037 & 0.469 & 0.468 & & 0 & 0 & 0.500 & 0.500\\ 

$|D\rangle$ & $0.1$ & & 2840 & 4220 & 9664 & 32592 & &0.058 & 0.086 & 0.196 & 0.661 & &0.050 & 0.050 & 0.167 & 0.733\\ 

 & $0.33$ && 7512 & 9496 & 1512 & 30916 && 0.152 & 0.192 & 0.031 & 0.625 && 0.167 & 0.167 & 0 & 0.667\\ \hline

 & $0$ & & 22664 & 23388 & 1140 & 1112 & & 0.469 & 0.484 & 0.024 & 0.023 & & 0.500 & 0.500 & 0 & 0\\ 

$|A\rangle$ & $0.1$ & & 8480 & 34492 & 4088 & 2052 & &0.173 & 0.702 & 0.083 & 0.042 & &0.167 & 0.733 & 0.050 & 0.050\\ 

 & $0.33$ && 1096 & 32360 & 9384 & 6564 && 0.022 & 0.655 & 0.19 & 0.133 && 0 & 0.667 & 0.167 & 0.167\\

\hline\hline
\end{tabular}
\caption{Data samples, estimated probabilities, and theoretical values for $D$ and $A$ inputs with Bob using the same basis as Alice, and for predicted error probabilities $P_E = 0, 0.1$, and 0.33.  $|0\rangle|1\rangle$ corresponds to Bob's measuring $|D\rangle$ and Eve's measuring $|1\rangle_{\rm T}$.   Column~1 shows the state Alice sent and  column~2 shows the predicted error probability $P_E$. ``Coincidence" columns show coincidence counts over a 40-s interval. ``Estimated" columns show the measured coincidence counts normalized by the total counts of all four detectors, and ``Expected" shows the theoretical values under ideal operating conditions.}
\label{TAB:DATA}
\end{table*}

Figure~\ref{QUANTUM_CIRCUIT} shows the quantum circuit diagram of our SPTQ implementation of the FPB probe.  We start with a pair of polarization-entangled photons in the singlet state.  Photon 1 is used as a trigger to herald photon 2 as a single-photon pulse for the BB84 protocol.  A {\sc swap} operation applied to photon 2 exchanges its polarization and momentum qubits so that the polarization of photon 1 and the momentum of photon 2 are now entangled in a singlet state.  Eve encodes her probe qubit in the momentum state of photon 2 by projecting photon 1 along an appropriate polarization state set by a polarization rotation ${\rm R_E}$.  The polarization state of photon 2 after the {\sc swap} gate is Alice's qubit, which is set by rotation ${\rm R_A}$.  Similarly, Bob's polarization analysis of Alice's qubit is set by ${\rm R_B}$.  The {\sc cnot} gate in Fig.~\ref{QUANTUM_CIRCUIT} is preceded by a $-\pi/8$ rotation and followed by a $+\pi/8$ rotation because the basis for the {\sc cnot}'s control qubit is rotated by $\pi/8$ from the BB84 bases, as noted in Fig.~\ref{Basis}.  The {\sc cnot} gate that Eve employs is a polarization-controlled {\sc not} ({\sc p-cnot}) gate that uses the polarization qubit as the control and the momentum qubit of the same photon as the target.  We have previously demonstrated such a gate in a polarization Sagnac interferometer with an embedded dove prism \cite{SPTQ}.  We have also demonstrated  the  {\sc swap} operation \cite{SWAP} by cascading three {\sc cnot} gates: a 
momentum-controlled {\sc not} ({\sc m-cnot}), a {\sc p-cnot}, and another  
{\sc m-cnot}.

Figure \ref{EXPSETUP} shows our experimental setup for implementing the quantum circuit.  We used a bidirectionally pumped Sagnac interferometric down-conversion source \cite{SAGNAC} with a periodically poled KTiOPO$_4$ (PPKTP) crystal to generate polarization-entangled photons at 810 nm in the singlet state.  The measured flux was $\sim$700 pairs/s per mW of pump in a 1 nm bandwidth at $\sim$99.45\% visibility quantum interference.  The collimated output beam had a beam waist of $w_0=0.53$ mm. In a collimated configuration, the momentum state of a photon is the same as the spatial orientation of the beam, which we use the right--left ($R$--$L$) basis in this experiment.  The $L$ and $R$ beams were separated by $\sim$2 mm.

For each photon pair, photon 1 is used to herald the arrival of photon 2 and also to remotely control the momentum qubit of photon 2 by postselection.  ${\rm R_E}$ polarization rotation by Eve was implemented using a quarter-wave plate (QWP) Q1 and a half-wave plate (HWP) H1, followed by single-photon detection (D1) through a polarizing beam splitter (PBS) P1 along $H$.  Q1 was used to compensate an intrinsic phase shift $\xi$ imposed by the {\sc swap} gate on the target-qubit basis  $\{|0\rangle_{\rm T},|1\rangle_{\rm T}\}$.  We have independently measured $\xi \simeq 88^\circ$.  Therefore, the ${\rm R_E}$ operation prepared the momentum qubit in $|T'_{\rm in}\rangle$ with a Q1-imposed phase shift $\xi$ 
\begin{equation}
|T'_{\rm in}\rangle \equiv \cos\theta_{\rm in}|0\rangle_{\rm T}+e^{i \xi}\sin\theta_{\rm in}|1\rangle_{\rm T}\,. \label{eq:T'_in}
\end{equation}
The extra phase shift of the {\sc swap} gate would bring Eve's probe qubit to be in $|T_{\rm in}\rangle$ of Eq.~\ref{eq:T_in}. 

After the {\sc swap} gate, ${\rm R_A}$ and ${\rm R_{-\pi/8}}$ were combined in a single operation.  The {\sc p-cnot} gate had the same phase shift problem as the {\sc swap} gate, so we used a HWP (H2) and a QWP (Q2) to compensate for  this phase shift and to impose the required rotation.  After H2 and Q2 Alice's qubit becomes 
\begin{equation}
|\Psi_{\rm A}\rangle \equiv \cos\theta_{\rm A}|0\rangle_{\rm C}+e^{i \chi}\sin\theta_{\rm A}|1\rangle_{\rm C}, \label{eq:Psi_A}
\end{equation}
where $\chi$ ($\simeq 98^\circ$) is the compensating phase shift and $\theta_{\rm A}$ is $-22.5^\circ, 22.5^\circ, 67.5^\circ$, or $112.5^\circ$ for $|H\rangle, |D\rangle, |V\rangle$ or $|A\rangle$, respectively, as shown in Fig.~\ref{Basis}(a).  Similarly we combined ${\rm R_{\pi/8}}$ and ${\rm R_B}$ into a single HWP (H3) in Fig.~\ref{EXPSETUP} and a PBS (P2) was used by Bob to analyze the polarization of Alice's qubit. 

Eve measured her qubit by a projective measurement along the 
$|0\rangle_{\rm T}$--$|1\rangle_{\rm T}$ (spatially, $R$--$L$) basis.  A HWP (H4/H5) was placed in the $R$ or $L$ beam path, as indicated in Fig.~\ref{EXPSETUP}, so that the $R$ and $L$ beams would be distinguished by their orthogonal polarizations.  This polarization tagging simplified their measurements by a PBS (P3/P4) and single-photon detectors.  The four detectors uniquely identified the two qubits of photon 2.  For example, D2 (D3) indicates $R$ ($L$) for Eve's qubit, and either D2 or D3 suggests $H$ polarization after Bob's analyzer.  Therefore, in our physical simulation, the joint measurement by Bob and Eve yields Bob's polarization information and Eve's momentum information.

\begin{figure}[t]
\center{\includegraphics[width=0.85\linewidth]{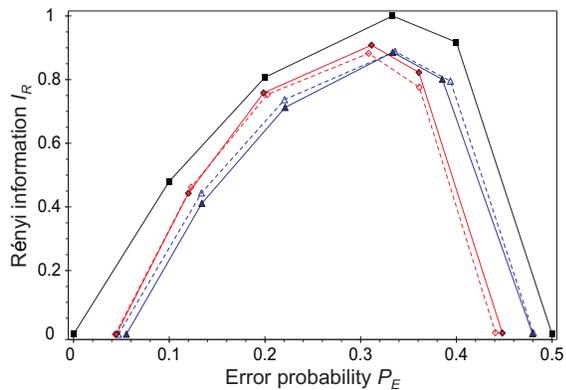}}
\caption{(Color online) Eve's R{\'e}nyi information $I_R$ about Bob's error-free sifted bits as a function of error probability $P_E$ that her eavesdropping creates. Solid squares, solid line: theoretical curve from Eq.~(\ref{eq:I_R}).  Solid diamonds, solid line: measured values for $H$--$V$ basis. Solid triangles, solid line: measured values for $D$--$A$  basis.  Open diamonds, dashed line: fitted curve with error model for 
$H$--$V$ basis.  Open triangles, dashed line: fitted curve with error model for $D$--$A$ basis.}
\label{RENYI_INFO}
\end{figure}

In data collection, we measured coincidences between D1 and one of the detectors for photon 2. Table \ref{TAB:DATA} shows two data sets for Alice's input of $D$ and $A$ polarizations and compares with the expected values for the ideal case. From the raw data, we calculate the R{\'e}nyi information $I_R$ based on Eq.~(\ref{def_I_R}), and Fig.~\ref{RENYI_INFO} plots $I_R$ as a function of the error probability $P_E$ for the ideal case (solid squares, solid line), for the measured values with inputs in the $H$--$V$ basis (solid diamonds, solid line), and for the measured values with inputs in the 
$D$--$A$ basis (solid triangles, solid line).  We note that no background counts were subtracted and the coincidence window was $\sim$1 ns.

In the ideal case with $P_E = 0$, Eve gets no information, $I_R = 0$, and Alice and Bob have no error bits.  However, due to experimental errors such as imperfect gate fidelities, we measured $\sim$5\%  of the sifted bits had errors.  For $P_E = 1/3$, Eve obtains perfect information, $I_R = 1$, but in our experiment, Eve gained a maximum $I_R = 0.9$ or, on average, Eve gained 95\% of the correct information about Bob's error-free sifted bits. 

To understand the errors involved in the experiment, we model our experimental setup with some non-ideal parameters. We assume that the phases $\xi$ in Eq.~(\ref{eq:T'_in}) and $\chi$ in Eq.~(\ref{eq:Psi_A}) could be inaccurate, and similarly for the setting of $\theta_{\rm A}$ in Eq.~(\ref{eq:Psi_A}) that might be caused by the wave plates. We also model the unitary {\sc p-cnot} gate  as
\begin{equation}
 \left( \begin{array}{cccc}
\cos\alpha & i e^{-i \delta}\sin\alpha & 0 &0 \\
i e^{i \delta}\sin\alpha & \cos\alpha & 0 & 0 \\
0 & 0 & -i e^{i \delta}\sin\alpha & \cos\alpha \\
0 & 0 & \cos\alpha & -i e^{-i \delta}\sin\alpha 
  \end{array} 
 \right),
\end{equation}
where $\alpha=0$ and $\delta=0$ for an ideal {\sc p-cnot} gate.
Finally we assume that Bob's HWP (H3) setting of $\theta_B$ was not perfect such that 
\begin{eqnarray}
|H\rangle \rightarrow \cos\theta_B |0\rangle_{\rm C}-\sin\theta_B|1\rangle_{\rm C}\,,\\
|V\rangle \rightarrow \sin\theta_B |0\rangle_{\rm C}+\cos\theta_B|1\rangle_{\rm C}\,,
\end{eqnarray}
where $\theta_B$ should equal $22.5^\circ$ ($-22.5^\circ$) in the $H$--$V$ ($D$--$A$) basis. 

We fit the data by minimizing the differences between 96 measurements and the calculated numbers based on this error model.  The fitting results show that $\Delta\xi\simeq3^\circ$, $\Delta\chi\simeq-11^\circ$, $\Delta\theta_A (H,D,V,A)=\{3.2^\circ, 0.9^\circ, -0.7^\circ, -2.3^\circ \}$, $\alpha=12.3^\circ$, $\delta=3.6^\circ$, $\Delta\theta_B (H/V,D/A)=\{-1.8^\circ,0^\circ\}$.    As expected, the phase errors are relatively small and those associated with $\theta_A$ and $\theta_B$ are within the resolution of the rotating mounts housing the wave plates.  The non-zero $\alpha$ also agrees with the measured classical visibility of 94\% for the {\sc p-cnot} gate.  We plot the fitted $I_R$ based on this model in Fig.~\ref{RENYI_INFO} for the $H$--$V$ basis (open diamonds, dashed line) and the $D$--$A$ basis (open triangles, dashed line). 

In summary, we have demonstrated experimentally the first complete physical simulation of the entangling-probe attack, showing that Eve can gain R{\'e}nyi information of up to 0.9 under realistic operating conditions, including a {\sc cnot} gate that does not have an ultrahigh fidelity.  Our results suggest the possible amount of information gain by Eve with current technology and the need to evaluate the required level of privacy amplification. 

This work was supported by BBN Technologies under the DARPA QuIST program.

\end{document}